\documentclass[iop,superscriptaddress]{emulateapj}
\pdfoutput=1
\usepackage{amsmath,epsfig,url,natbib}
\usepackage{graphicx,epstopdf,float,color,array}
\usepackage{rotating,multirow}
\journalinfo{\@Accepted for Publication in the Astrophysical Journal}
\shortauthors{Hemmati et al.}
\begin{document}
\title{Nebular and Stellar Dust Extinction across the Disk of
emission-line galaxies on small (kpc) scales}
\author{
Shoubaneh Hemmati\altaffilmark{1},
Bahram Mobasher \altaffilmark{1},
Behnam Darvish \altaffilmark{1},
Hooshang Nayyeri \altaffilmark{2},
David Sobral \altaffilmark{3,4},
Sarah Miller \altaffilmark{2}
}

\email{shemm001@ucr.edu}
\altaffiltext{1}{University of California, Riverside, CA 92512}
\altaffiltext{2}{University of California, Irvine, CA 92697}
\altaffiltext{3}{Universidade de Lisboa, PT1349-018 Lisbon, Portugal}
\altaffiltext{4}{Leiden Observatory, NL-2300 RA Leiden, The Netherlands}

\begin{abstract}

We investigate resolved kpc-scale stellar and nebular dust distribution in eight star-forming
galaxies at $z\sim0.4$ in the GOODS fields. This is to get a
  better understanding of the effect of dust attenuation on
  measurements of physical properties and its variation with
  redshift. Constructing the observed Spectral Energy Distributions (SEDs) per pixel, based on seven bands
photometric data from HST/ACS and WFC3, we performed pixel-by-pixel SED fits to population synthesis models and estimated small-scale distribution of stellar dust extinction. We use $\rm H\alpha / H\beta$
nebular emission line ratios from Keck/DEIMOS high resolution spectra at each
spatial resolution element to measure the amount of attenuation faced
by ionized gas at different radii from the center of galaxies. We find
a good agreement between the integrated and median of resolved color
excess measurements in our galaxies. The ratio of integrated nebular to stellar
dust extinction is always greater than unity, but does not show any trend
with stellar mass or star formation rate. We find that inclination
plays an important role in the variation of the nebular to stellar
excess ratio. The stellar color excess profiles are found to have
higher values at the center compared to outer parts of the
disk. However, for lower mass galaxies, a similar trend is not found
for the nebular color excess. We find that the nebular color excess
increases with stellar mass surface density. This explains the absence
of radial trend in the nebular color excess in lower mass
galaxies which lack a large radial variation of stellar mass
  surface density. Using standard conversions of star formation rate
surface density to gas mass surface density, and the relation between dust
mass surface density and color excess, we find no significant
variation in the dust to gas ratio in regions with high gas mass
surface densities, over the scales probed in this study.

\end{abstract}
\keywords{galaxies: evolution --- galaxies: fundamental parameters --- galaxies: kinematics and dynamics --- galaxies: spiral}

\section{Introduction}

The existence of interstellar dust was suggested about a century
ago, even before the great debate about the nature of galaxies
(\citealt{Curtis1918}). The presence of dust in galaxies was later
established firmly, not only from dimming and reddening of light but
also from dust scattering and Far Infrared (FIR) continuum emission. Dust can absorb more than half of the Ultraviolet (UV) and optical radiation
budget of the Universe (e.g. \citealt{Calzetti2001}). Most of the radiation from star
formation in galaxies is emitted in the UV and optical, the wavelength range most susceptible to dust extinction and where most of the observations are taken. Therefore, without a thorough
understanding of the attenuation of light of galaxies (coming from
both stars and nebulae) by dust, interpretation of their physical
properties (e.g. star formation rate, stellar age and stellar mass-to-light ratio) will not be accurate. 

A lot of progress has been made in our understanding of the attenuation of
light by dust. It is now well established that the amount of extinction in
galaxies is wavelength dependent. Variation of dust extinction as a
function of wavelength, or the so called extinction curve, is studied
from different observations of nearby galaxies and the Milky Way
(e.g. \citealt{Prevot1984}, \citealt{Cardelli1989}, \citealt{Calzetti1994},
\citealt{Gordon2003}). In the extinction curve, there is information
about chemical composition and sizes of dust grains. The smoothness of the
FIR part of the extinction curve suggests that a variety of grain sizes
exist. While the overall shape of the extinction curves measured from
these local galaxies agrees, especially towards the infrared, there
are some differences in the slopes, the normalizations, and the
presence of the 2175 \AA \ bump (e.g. \citealt{Stecher1965}, \citealt{Calzetti2001}, \citealt{Reddy2015}, \citealt{Scoville2015}). These differences are attributed to different metallicities and dust-to-gas ratios
as well as differences in the composition of the dust grains (e.g. \citealt{Calzetti1994}, \citealt{Reddy2015}).

Local extinction and attenuation curves are often used to correct for
dust attenuation at intermediate and high redshifts. Recently, the study of dust attenuation at higher redshifts has become accessible by infrared surveys. Studies at $z \sim 1-2$
(e.g. \citealt{Scoville2015}, \citealt{Reddy2015}) have found
attenuation curves very similar to the commonly used
\cite{Calzetti2000} attenuation curve derived from nearby galaxies. However, other
studies reported poor fits from nearby curves to high redshift
galaxies and found strong spectral dependence in the attenuation curve
in their sample (e.g. \citealt{Kriek2013}), suggesting different
star-dust geometry, dust grain properties or both. 

Comparison of attenuation towards nebular star forming regions inside
galaxies with that of the stellar continuum hints about the geometry
of dust relative to stars. Studies of local galaxies have found higher attenuation
towards the nebular regions compared to the integrated dust from
stellar continuum (e.g. \citealt{Calzetti2000},
\citealt{Moustakas2006}, \citealt{Wild2011}, \citealt{Kreckel2013}). This is consistent with the picture from
the radiative transfer model by \cite{Charlot2000} in which recombination
lines generated in the HII regions, the birth clouds of the most massive O
stars, face an extra amount of attenuation by dust. However, at higher
redshifts the picture is not as clear. By comparing star formation rates
from different diagnostics, it was found that (e.g. \citealt{Erb2006},
\citealt{Reddy2010}, \citealt{Garn2010}, \citealt{Shivaei2015},
\citealt{Oteo2015}) the best agreement between the SFRs is met with no
extra color excess towards the nebular regions. Other studies have
found quite the opposite, with higher attenuation needed towards the nebular regions
(e.g. \citealt{Forster2009}, \citealt{wuyts2011}, \citealt{Ly2012}
\citealt{Price2014}, Darvish et al. (in prep)). Using a large sample
of emission line galaxies from the MOSFIRE Deep Evolution Field (MOSDEF) survey
(\citealt{Kriek2015}), \cite{Reddy2015} demonstrated that the ratio of nebular to stellar attenuation is a
function of the star formation and specific star formation rates of galaxies. However, there
is a huge scatter in the relation. A detailed analysis of spatially
resolved colors is needed to understand the source of this scatter.

The amount of nebular attenuation by dust is also shown to correlate with
physical properties of galaxies, such as luminosity, Star Formation Rate (SFR),
mass and metallicity (e.g. \citealt{Wang1996}, \citealt{Sullivan2001},
\citealt{Pannella2009}, \citealt{Asari2007}). \cite{Garn2010}
investigated these different dependencies and found stellar mass to be the
best parameter predicting the amount of dust extinction for $z\sim0.1$
galaxies from the Sloan Digital Sky Survey (SDSS DR7;
\citealt{Abazajian2009}). This result was later confirmed by studies
at high redshift galaxies (e.g. \citealt{Sobral2012},
\citealt{Dominguez2013}, \citealt{Ibar2013}). It is however not clear
whether these relations hold at smaller scales inside galaxies.

High spatial resolution observations of galaxies in the local
Universe enabled the derivation of well calibrated relations among
the physical properties of galaxies (such as stellar mass, dust mass,
or SFR) at $z\sim 0$ (e.g. \citealt{Calzetti2000},
\citealt{Kennicutt2007}, \citealt{Kreckel2013},
\citealt{Boquien2011}). However until recently, at intermediate and
high redshifts, studies of resolved (kpc-scale) properties of galaxies
was not possible. High resolution multi-waveband surveys using the
Hubble Space Telescope such as Cosmic Assembly Near-infrared Deep
Legacy Survey (CANDELS; \citealt{Grogin2011}, \citealt{Koekemoer2011}) have enabled such
studies through optical and near-infrared photometric observations to
$z\sim 2$ (e.g. \citealt{Wuyts2012}, \citealt{Hemmati2014},
\citealt{Guo2015}). In these studies, resolved physical properties of
galaxies were measured by fitting the observed Spectral Energy
Distribution (SED) at each pixel to the theoretical stellar synthesis
models. Therefore, the uncertainties inherent to SED fitting will be
inevitable using only photometric data. Spectroscopic data from grism observations by surveys such as 3D-HST have
improved measurements of resolved physical properties (such as SFR
surface densities).

Using a sample of massive galaxies at z~1 in the CANDELS and 3D-HST,
\cite{Wuyts2013} showed that an extra amount of attenuation is needed
for nebular gas for the integrated H$\alpha$ SFR to agree with the SED
inferred SFR. However, one caveat of such studies is the poor spectral
resolution of grism observations, which can not resolve the NII and H$\alpha$ emission lines.
The contribution of the NII to the H$\alpha$+NII emission depends on
the ionization radiation and metallicity of the galaxy and hence a
constant ratio assumption could affect the inferred SFRs measured. 
 The advent of Adaptive Optics
(AO) aided Integral Field Spectrographs (IFS), enabled spatially
resolved spectroscopic observations of intermediate redshift galaxies
(e.g. \citealt{Forster2009}, \citealt{Genzel2010}, \citealt{Swinbank2012}). While obtaining statistical samples of galaxies with IFS has become possible over the last years (e.g. \citealt{Sobral2013}, \citealt{Wisnioski2015}), the same is not yet true for AO aided high resolution IFS observations.

In this work, we combine high resolution photometric data from CANDELS with
complementary high spatial and spectral resolution spectroscopic data
from Keck/DEIMOS observations to study dust distribution in a sample
of emission-line galaxies. We demonstrate the usefulness of the
technique on a small sample. We measure the stellar continuum and
ionized gas dust extinction along the major axis of disk
galaxies. Stellar Continuum and the ionized gas extinctions are
measured from resolved SED-fitting per pixel (\citealt{Hemmati2014}) and the Balmer decrements from Keck/DEIMOS
spectra, respectively. We investigate how integrated dust measurements in
galaxies compare to the spatial variation of dust along the disks.

The structure of this paper is as follows. \S 2 presents the sample
selection and data. In \S 3 and \S 4 we describe measurements from
photometric and spectroscopic data, respectively. We present our
results in \S 5 and in \S 6 we finish with a summary and discussion. Throughout this paper all
magnitudes are expressed in AB system \citep{Oke1983} and we use
standard cosmology with $H_{0}=70 \rm \:kms^{-1}
Mpc^{-1}$, $\Omega_{M}=0.3$ and $\Omega_{\Lambda}=0.7$.

\section{Sample and Data}

The sample for this study consists of eight disky galaxies at
$z\sim0.4$ with median stellar mass of $\rm log(M_{*}/M{\odot})=9.6$
and median SFR of $\rm SFR(M_{\odot}/yr)= 10.0$. These galaxies are selected from a parent sample
described in detail in \cite{Bundy2005} and
\cite{Miller2011}. Here, we present the selection criteria
for the sample used in this study. 

\begin{figure*} []
\centering
\begin{tabular}{|c|c|}
\hline\\
\includegraphics[trim=3cm 2cm 1.5cm 0.8cm, clip,width=9cm]{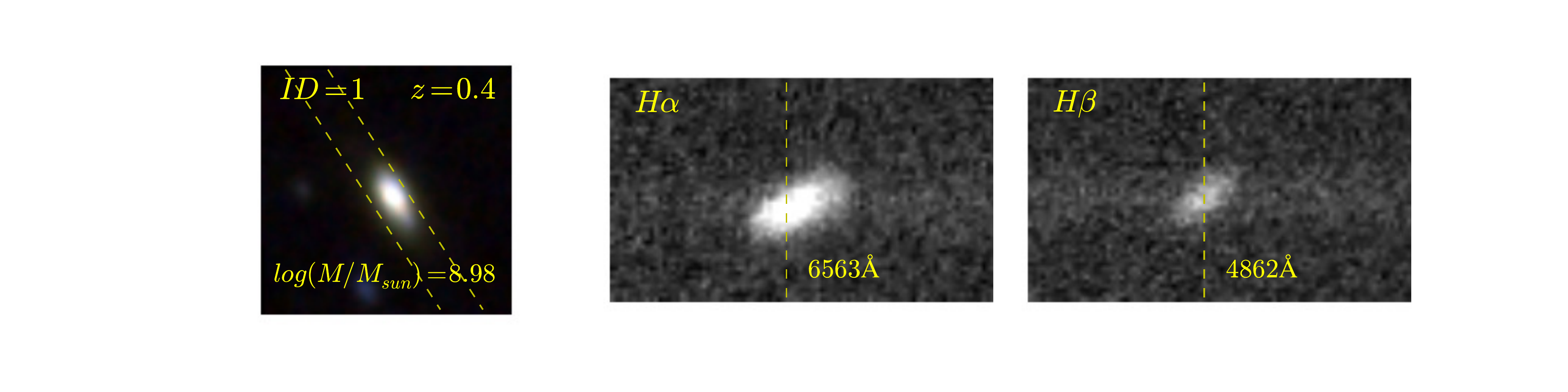}
\includegraphics[trim=3cm 2cm 1.5cm 0.8cm, clip,width=9cm]{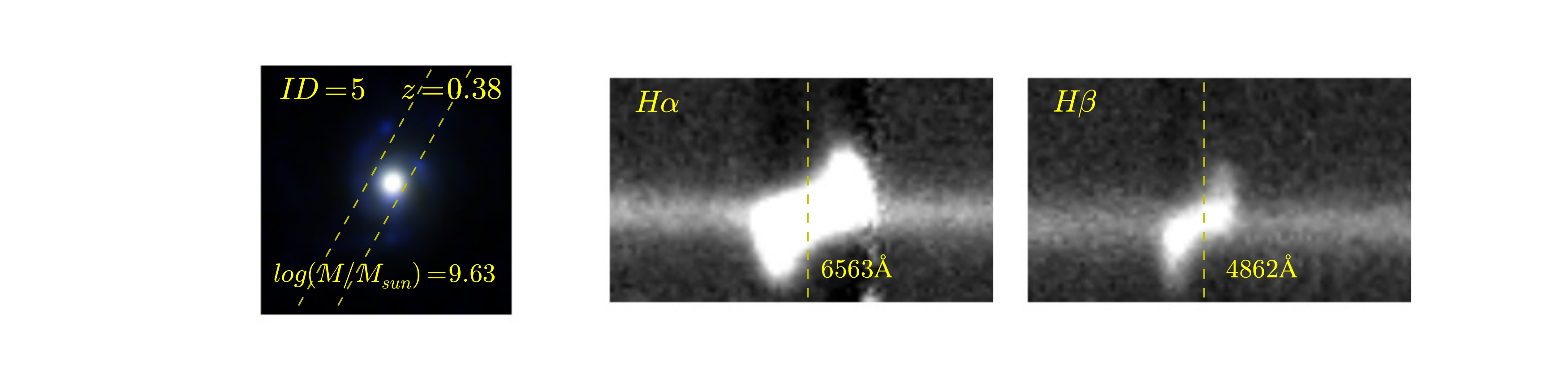}\\
\includegraphics[trim=3cm 2cm 1.5cm 0.8cm, clip,width=9cm]{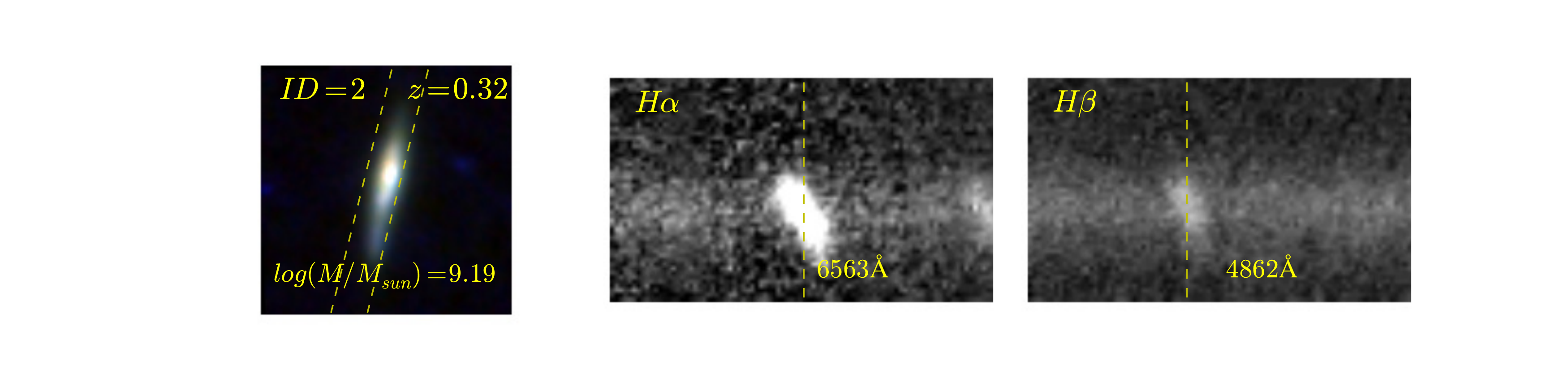}
\includegraphics[trim=3cm 2cm 1.5cm 0.8cm, clip,width=9cm]{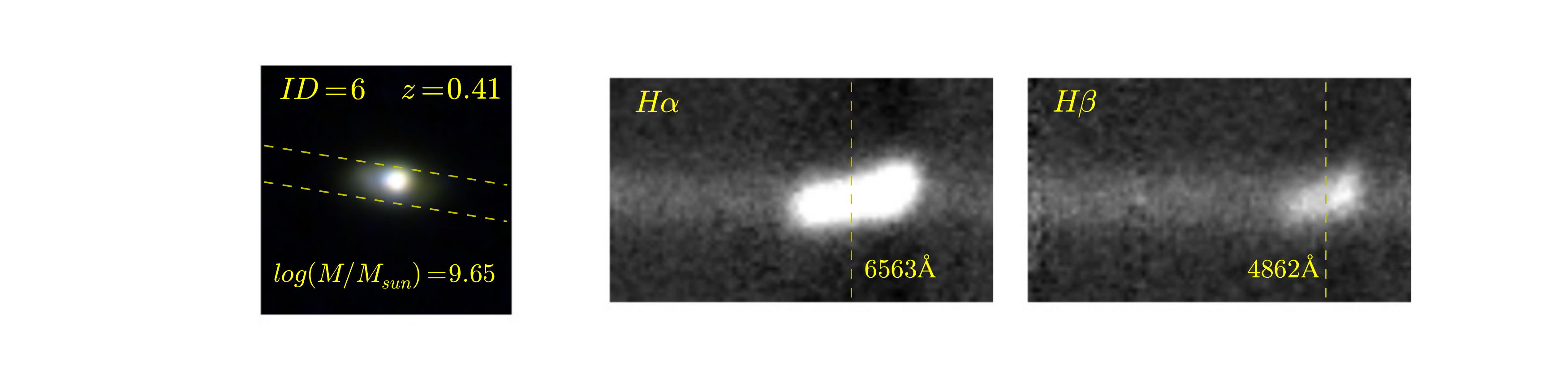}\\
\includegraphics[trim=3cm 2cm 1.5cm 0.8cm, clip,width=9cm]{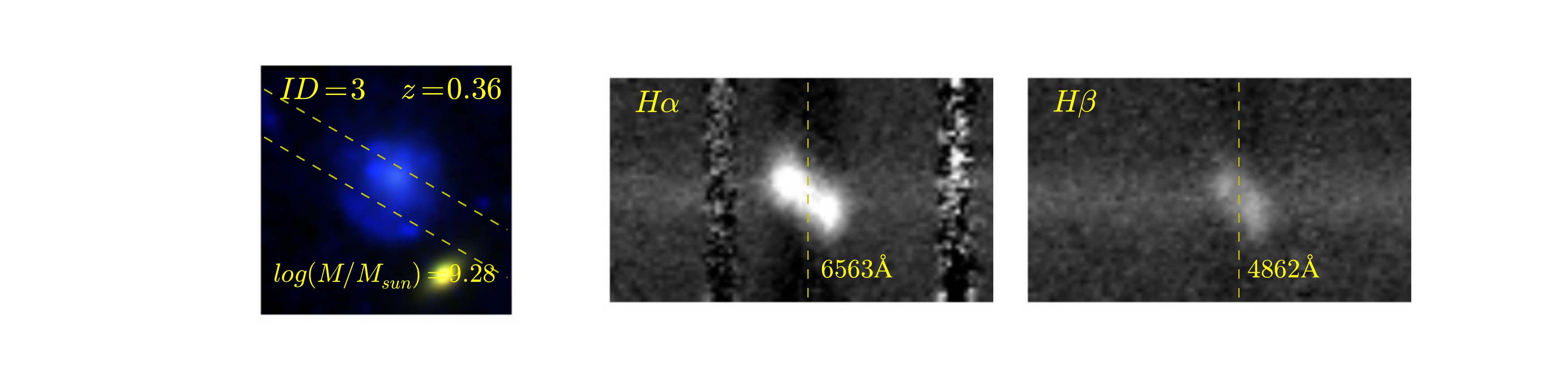}
\includegraphics[trim=3cm 2cm 1.5cm 0.8cm, clip,width=9cm]{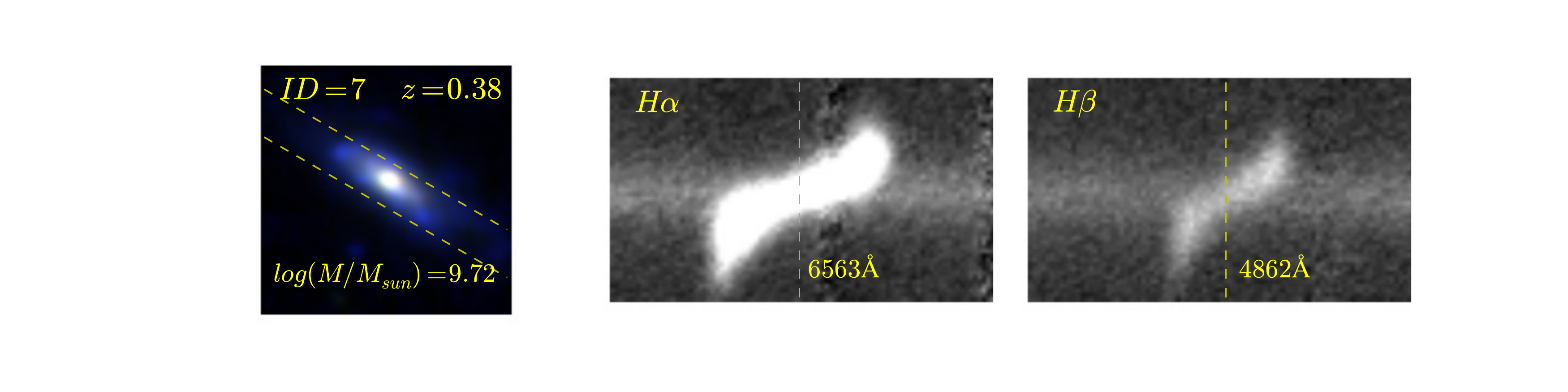}\\
\includegraphics[trim=3cm 2cm 1.5cm 0.8cm, clip,width=9cm]{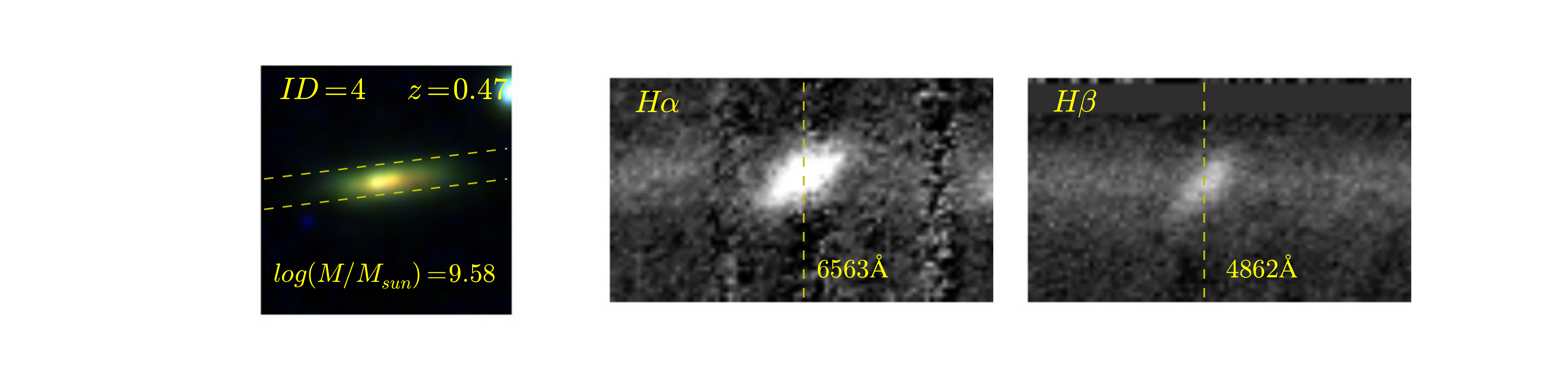}
\includegraphics[trim=3cm 2cm 1.5cm 0.8cm, clip,width=9cm]{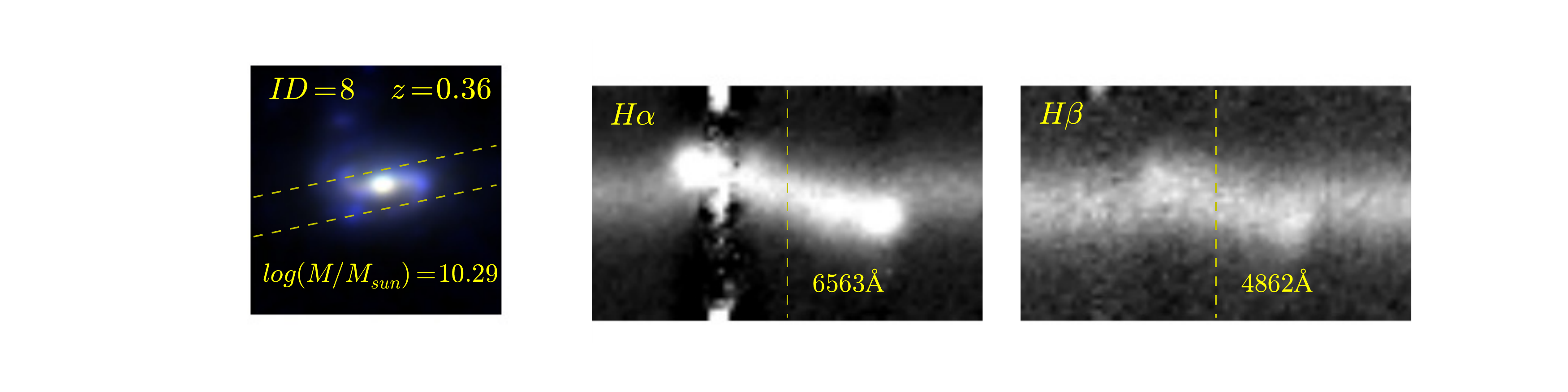}\\
\\
\hline
\end{tabular}
\caption{HST ($B_{F435W}$, $Z_{F850lp}$, $H_{F160W}$) stacked images of galaxies in the sample as well as their $H\alpha$ and $H\beta$ emission line cutouts from the DEIMOS 2D spectra, sorted based on stellar mass. The stacked images are $80$ pixels in each side ( $\sim 25$ kpc at $z \sim 0.4$). Yellow dashed lines on images of galaxies show DEIMOS 1'' slit. }
\end{figure*}

Galaxies in the parent sample were initially adapted from a
$Z_{F850LP}<22.5$ sample in the Great Observatories Origins Deep
Survey (GOODS) North and South fields (\citealt{Giavalisco2004}) in
the redshift (spectroscopic when available) range of $0.2<z<1.3$ and
were visually inspected to have prominent disks. A further magnitude
cut of $K_{s} < 22.2$ was applied to ensure reliable stellar mass
measurements. We then observed these galaxies with the DEIMOS (\citealt{Faber2003})
instrument on KECK with long ($\sim$ 6-8 hours) exposures. The $1200$ 
 $\rm l mm^{-1}$ grating and $1\arcsec$ slits with a central wavelength
of 7500 \AA \ were used, achieving FWHM spectral resolution of $\sim 1.7$
\AA. The PA of slits are almost aligned with the
galaxies major axis. This was to measure rotation curves for the
  galaxies in the parent sample (\citealt{Miller2011}). Out of
  the galaxies with extended line emission, only eight have
both H$\alpha$ and H$\beta$ covered in the DEIMOS spectra which are not
contaminated by OH sky emission lines. These galaxies are selected for
the analysis of this pilot study. The photometric data in this study comes from high resolution Hubble
Space Telescope (HST) optical and near infrared images taken by the
Advanced Camera for Surveys (ACS) and Wide Field Camera3 (WFC3) as
part of the CANDELS. We use HST/ACS observations in the F435W, F606W,
F775W and F850LP (hereafter $B_{F435W}$, $V_{F606W}$, $I_{F775W}$ and
$Z_{F850lp}$) and HST/WFC3 observations in the F105W, F125W and F160W
(hereafter $Y_{F105W}$, $J_{F125W}$ and $H_{F160W}$) filters. The ACS
images have been multi-drizzled to the WFC3 pixel scale of 0\farcs06
\citep{Koekemoer2011}. Figure 1 shows the stacked HST ($B_{F435W}$,
$Z_{F850lp}$ and $H_{F160W}$) images as well as cutouts of H$\alpha$
and H$\beta$ emission lines from the DEIMOS 2D spectra of all galaxies
in the sample.

\section{Photometric measurements}

Using the photometric data, we have made 2D maps of physical parameters
of galaxies (such as stellar mass and star formation rate surface
density, age and extinction) in the sample by measuring the observed
SED for individual pixels and fitting them to template SEDs. The
method and the uncertainties in the estimated parameters are discussed
in \cite{Hemmati2014}. Here, we briefly
explain the method for producing 2D maps. We use the 2D maps to
produce 1D profiles along the disk of the galaxy. These profiles will
then be used to directly combine/compare with spectroscopic data from
DEIMOS.

\begin{table*}
\centering
\caption{Integrated Properties of Galaxies in the Sample}
\begin{tabular}{ ccccccccc}
\hline
\hline
ID & RA & DEC & Spec z & sin (i\footnotemark[1]) & $Log(M_{*}/M_{\odot})$ \footnotemark[2]& $SFR (M_{\odot}yr^{-1})$ \footnotemark[2]& $E(B-V)_{star}$ \footnotemark[2]&$E(B-V)_{nebular}$\footnotemark[3]\\ \hline
1 & 53.0728469& -27.7376055 & 0.40 &0.92 & $8.98 \pm 0.11$ & $1.74_{-1.41}^{+4.60}$ & 0.10 & $0.52\pm 0.26$ \\ 
2 & 189.2838000& 62.3300500& 0.32&1.00 & $9.19 \pm 0.12$& $1.68_{-1.43}^{+5.07}$ & 0.15 & $0.59\pm 0.32$ \\ 
3 & 189.2643000& 62.2753800 &0.36 &0.67& $9.28 \pm 0.11$& $3.31_{-2.48}^{+7.69}$ & 0.15 &$ 0.41\pm 0.27$ \\ 
4 & 189.2070900& 62.2203900&0.47& 0.98 & $9.58 \pm 0.30$& $17.53_{-13.89}^{+40.34}$ & 0.4&$0.91\pm 0.18$ \\ 
5 & 189.2692900& 62.2811400&0.38&0.71 & $9.63 \pm 0.22$& $13.83_{-8.45}^{+16.74}$ & 0.4 &$ 0.99\pm 0.06$ \\ 
6 & 189.3240100& 62.2443400&0.41 &0.87& $9.65 \pm 0.13$& $6.36_{-4.58}^{+9.13}$ & 0.4 & $0.81\pm 0.13$ \\ 
7 & 189.1654100& 62.2574100&0.38&0.97 & $9.72 \pm 0.20$& $29.28_{-18.88}^{+40.50}$ & 0.45 &$0.86\pm 0.15$ \\ 
8 & 189.2497900& 62.2877800&0.36&0.88 & $10.29 \pm 0.10$& $21.17_{-8.47}^{+12.86}$& 0.5 &$1.22\pm0.21$ \\ 
\hline
\end{tabular}
\footnotetext[1]{Inclination}
\footnotetext[2]{From SED fitting}
\footnotetext[3]{From the Balmer Decrement}
\end{table*}

\begin{figure*} []
\centering
\includegraphics[trim=2cm 0cm 2cm 0cm, clip,width=0.9\textwidth]{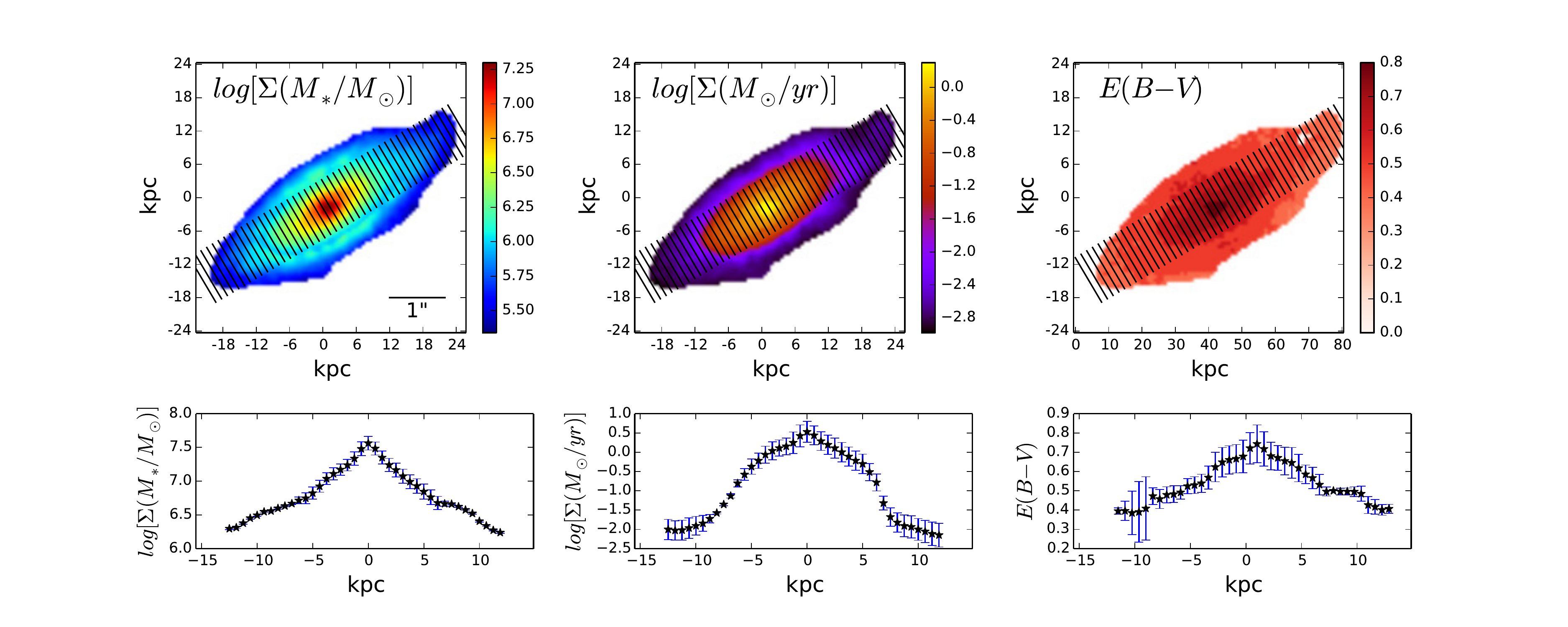}
\caption{ 2D maps from pixel-by-pixel SED fitting are on the first and 1D profiles along
  the major axis of the galaxy are on the second row.
Left to right, stellar mass surface density, SFR surface density and
$E(B-V)_{star}$ distribution of an example galaxy. The black solid
lines on top of 2D maps, represent the coverage and spatial resolution
of DEIMOS observation of this galaxy.}
\end{figure*}

We first make $80\times80$ pixel cutouts of galaxies in seven bands HST
science and rms error images. We PSF-match them to the resolution of the $H_{F160W}$
band. By multiplying the segmentation maps from SExtractor
(\citealt{Bertin1996}) to the PSF-matched cutouts we define the boundary of galaxy and remove any
surrounding objects. Using the LePhare code (\citealt{Arnouts1999};
\citealt{Ilbert2006}), we then fit the SED measured for each pixel with spectral
synthesis models to obtain the physical properties at that pixel (redshift is fixed to the spectroscopic redshift of the galaxy). 

The model library is built using BC03 (\citealt{BC03}) models,
Chabrier (\citealt{Chabrier2003}) Initial Mass Function (IMF), solar metallicity, declining star formation histories with a range of $\tau$
(including constant and bursts) and ages less than the age of the
Universe at the redshift of the galaxy in question. We use Calzetti starburst
(\citealt{Calzetti2000}) attenuation curve with a range of E(B-V) from
zero to one. We also include nebular emission lines in the fitting
procedure. LePhare code accounts for the contribution of emission lines
with a simple recipe based on the Kennicutt relations (\citealt{Kennicutt1998}). The
following lines are included in this treatment: Ly$\alpha$, H$\alpha$,
H$\beta$, [OII], [OIII]4959 and [OIII]5007 with different ratio with
respect to [OII] line as described in \citealt{Ilbert2008}. The
intensity of the lines are scaled according to the intrinsic UV
luminosity of the galaxy. The 2D maps of physical
parameters from the SED fitting output correspond to the median of the probability
distribution function marginalized over all other parameters. We use
$16\%$ lower and higher values from the Maximum Likelihood
analysis to measure the $1\sigma$ error for each parameter. Using
the same library and code, we also measure integrated properties of
galaxies by fitting the integrated light of all the pixels in the
defined boundary of the galaxy in the same seven ACS and WFC3
bands. Table 1 summarizes the properties of the host galaxies.

To be able to compare these measurements with their spectroscopic counterparts from Keck/DEIMOS, we need to convert 2D maps to 1D profiles. We combine
measurements of all pixels which correspond to one spatial resolution in
the DEIMOS spectra. In making the profiles, we only include pixels
that are covered by the DEIMOS slit. This is to avoid uncertain slit
loss corrections in the spectral measurements. Therefore, in the
direction perpendicular to the major axis of the galaxy, we bin all pixels
in the slit and bin again in the direction of major axis of the galaxy
to match the spatial pixel scale of DEIMOS (0.12\arcsec per
pixel). Figure 2 shows the 2D maps and 1D
profile of stellar mass, SFR and E(B-V) for one of the galaxies in the
sample.

\section{Spectral measurements}

\begin{figure*} []
\centering
\includegraphics[trim=0cm 0cm 0cm 0cm, clip,width=0.7\textwidth]{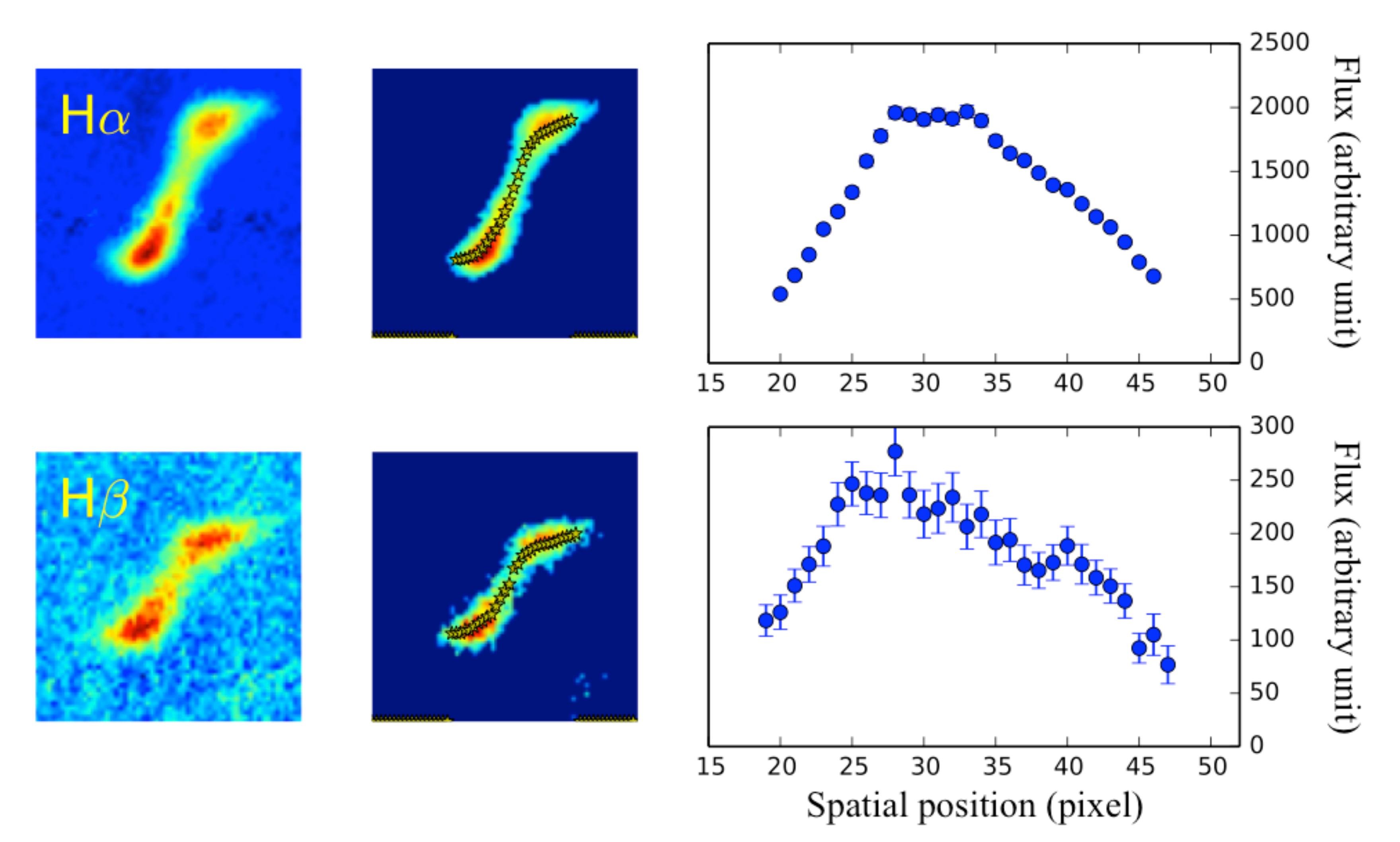}
\caption{Left to right, emission line cutout, thresholded emission and
total flux as a function of spatial position for H$\alpha$ (top) and
H$\beta$ (bottom) of an example galaxy. Yellow stars over plotted on
thresholded emission show peaks of Gaussians at each spatial position.}
\end{figure*}

We measure nebular dust attenuation along the disk of the galaxies
using the ratio of the first two Balmer transitions ($H\alpha$ and $H\beta$)
or the so-called Balmer decrement. Ratio of the luminosities of Balmer transitions
arising from HII regions around very massive stars is known to be the most
practical indicator measuring attenuation by dust in these
regions. In the absence of dust, the intrinsic theoretical H$\alpha$ to H$\beta$
line ratio, assuming Case B recombination, a
temperature of $T=10^4K$, and an electron density of
$n_e=10^2$ $\rm cm^{-3}$ is known ($F(H\alpha)/F(H\beta)=2.86$; \citealt{Osterbrock1989}). Any
deviation from this amount is indicative of the amount of dust
attenuation. We model the line emission at each spatial
resolution element of continuum subtracted DEIMOS spectra. We convert the H$\alpha$ to
H$\beta$ line ratio at each spatial resolution to an extinction
measure using:

\begin{equation}
E(B-V)_{nebular}=\rm \frac{-2.5}{k(H\beta)-k(H\alpha)}log_{10}(\frac{2.86}{H\alpha/H\beta})
\end{equation}

We use the Cardelli Galactic extinction curve (\citealt{Cardelli1989})
for measuring $\rm k(H\beta)$
and $\rm k(H\alpha)$. A line-of-sight attenuation curve such as Cardelli, is more
appropriate for recombination emission of compact HII regions compared
to more extended stellar continuum attenuation curves such as
\cite{Calzetti2000}. The latter would cause smaller $E(B-V)_{nebular}$ values by a factor of
0.9 (see \citealt{Reddy2015} for a more detailed discussion).

We model and subtract the continuum by first making cutouts of $\sim
100$ \AA \/ around the emission line (H$\alpha$ or H$\beta$ in this case). We mask
the wavelength range ($\sim 15-20$ \AA) where we have the emission in the cutout and bin
the rest of the spectra in the wavelength direction with bins of $\sim
5$ \AA. The median values in each wavelength bin are then
fitted with a Gaussian function in the spatial direction. We measure the continuum
under emission by interpolating these values over the masked
wavelength range and subtract this continuum from the cutout spectra
to have the continuum free emission line.

To model the line emission at each spatial resolution in the galaxy,
we make small cutouts covering only the emission from the continuum
subtracted spectra. To be able to trace the emission to furthest
points from the center and to avoid tracing and fitting the noise in
the background, we use the Otsu method (\citealt{Otsu1979}). Otsu
method is an unsupervised automatic thresholding algorithm which
reduces a grey level image to a binary image by finding an optimum
threshold to maximize the separability of the two classes. Here, using
this method we separate the emission from the noise in the background
by multiplying the binary image by the continuum subtracted cutout. We
fitted Gaussian functions to the emission at each spatial resolution
(in the wavelength/velocity direction) to measure the total flux from
that resolution element. In figure 3, we show H$\alpha$ and
H$\beta$ cutouts as well as the thresholded cutout and traced emission for
one example galaxy. We have centered emission
lines by the spectroscopic redshift, which comes from visually
aligning all the emission lines. However, in order to measure the ratio
of two lines we need to be more precise in centering the emissions. We
center each emission line by fitting the peak of modeled Gaussians
with an arctangent function (in the spatial direction) this takes
  care of velocity offsets due to rotation in the disk. We
  measure a correction factor from standard stars to account for flux calibration. This is an essential
  step  even though we are only using the ratio of emission lines, due to the large difference between wavelength
  of H$\alpha$ and H$\beta$ and the wavelength dependence
  of flux calibration. This is a factor of $\sim 1.2-1.5$ from
  $z=0.32-0.47$. The Balmer decrement is then measured along the major axis of galaxies by setting a minimum
signal-to-noise ratio (S/N) of 5 and 3 at each spatial
resolution for H$\alpha$ and H$\beta$ lines respectively and then
dividing the two. Nebular color excess ($E(B-V)_{nebular}$) along the disk
is then calculated by the conversion in equation 1. The uncertainty in
this ratio is calculated by the standard error propagation method
using the H$\alpha$ and H$\beta$ flux uncertainties at each spatial
position. In the resolved line measurement, we did not correct
for the Balmer absorption because the corrections are very small and
well within the uncertainty of the line measurements. Extinction
uncertainties range from 0.03-0.2 magnitude and its variation is
larger between galaxies compared to that of one single galaxy. 

We also measure the integrated or ``global'' Balmer decrement and nebular
color excess values for the whole galaxy by extracting the 1D spectra
and measuring line fluxes. We correct our measurements for the
underlying Balmer absorption using the stellar population model
fits. The corrections to the H$\alpha$ to H$\beta$ ratio are $\lesssim 4\%$. For simplicity,
we refer to our dust measurements at each spatial distance from the
center as resolved dust measurement. 

\section{Results}
\subsection{Color Excess, Stellar Continuum vs. ionized gas}

\begin{figure}[]
\includegraphics[trim=0cm 0cm 0cm 0cm, clip, width=0.51\textwidth]{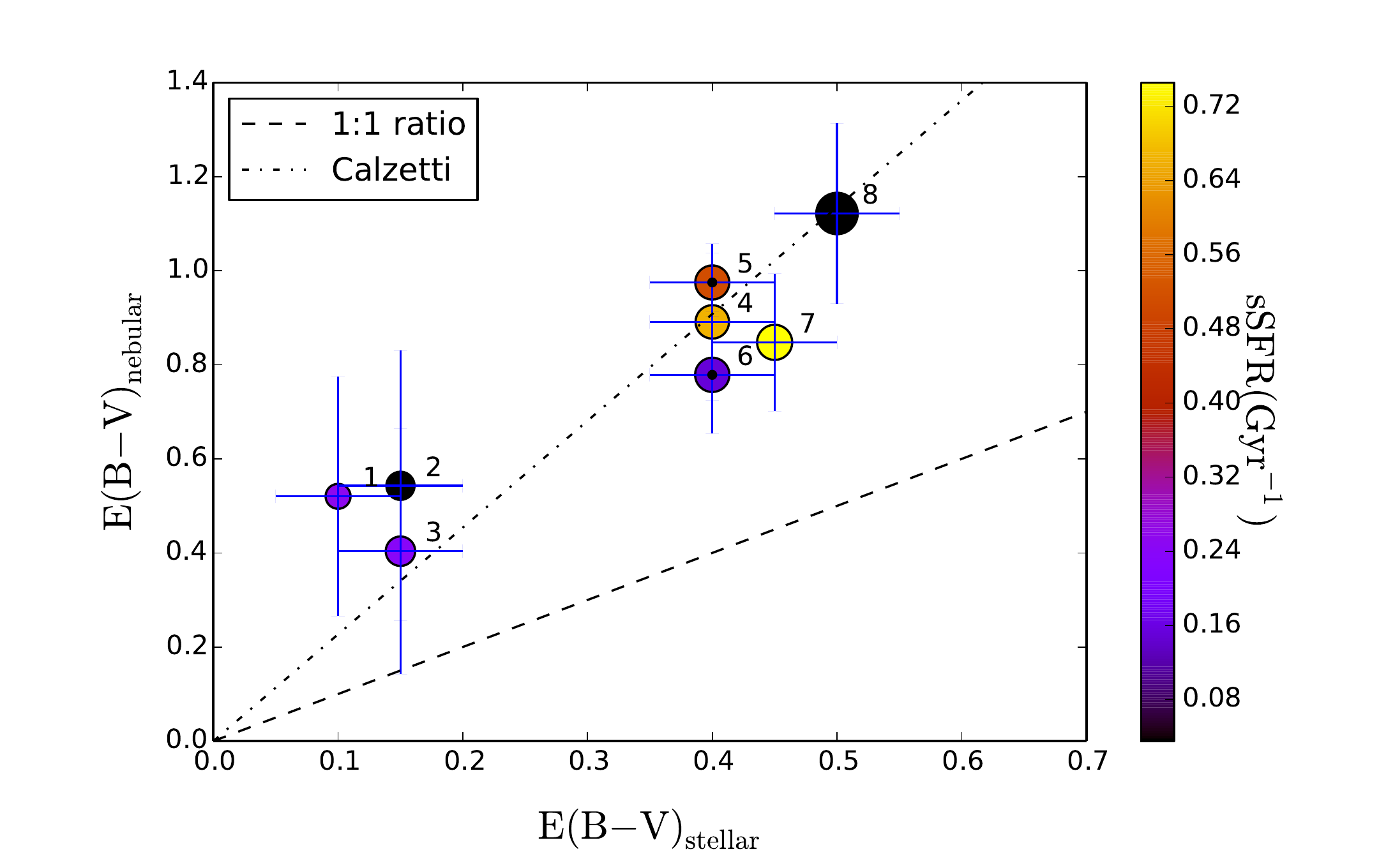}
\caption{Integrated nebular color excess of galaxies in the sample measured through the Balmer decrement compared to their stellar color excess measured through SED fitting. Nebular to stellar color excess ratio of 1 and 2.27 (\citealt{Calzetti2000}) are shown with dashed and dotted-dashed lines respectively. Galaxies are color coded based on their sSFR and symbol size increases with stellar mass.}
\end{figure}

\begin{figure*} []
\centering
\begin{tabular}{cc}
\includegraphics[trim=0cm 0.1cm 0.5cm 0.5cm, clip,width=7.2cm]{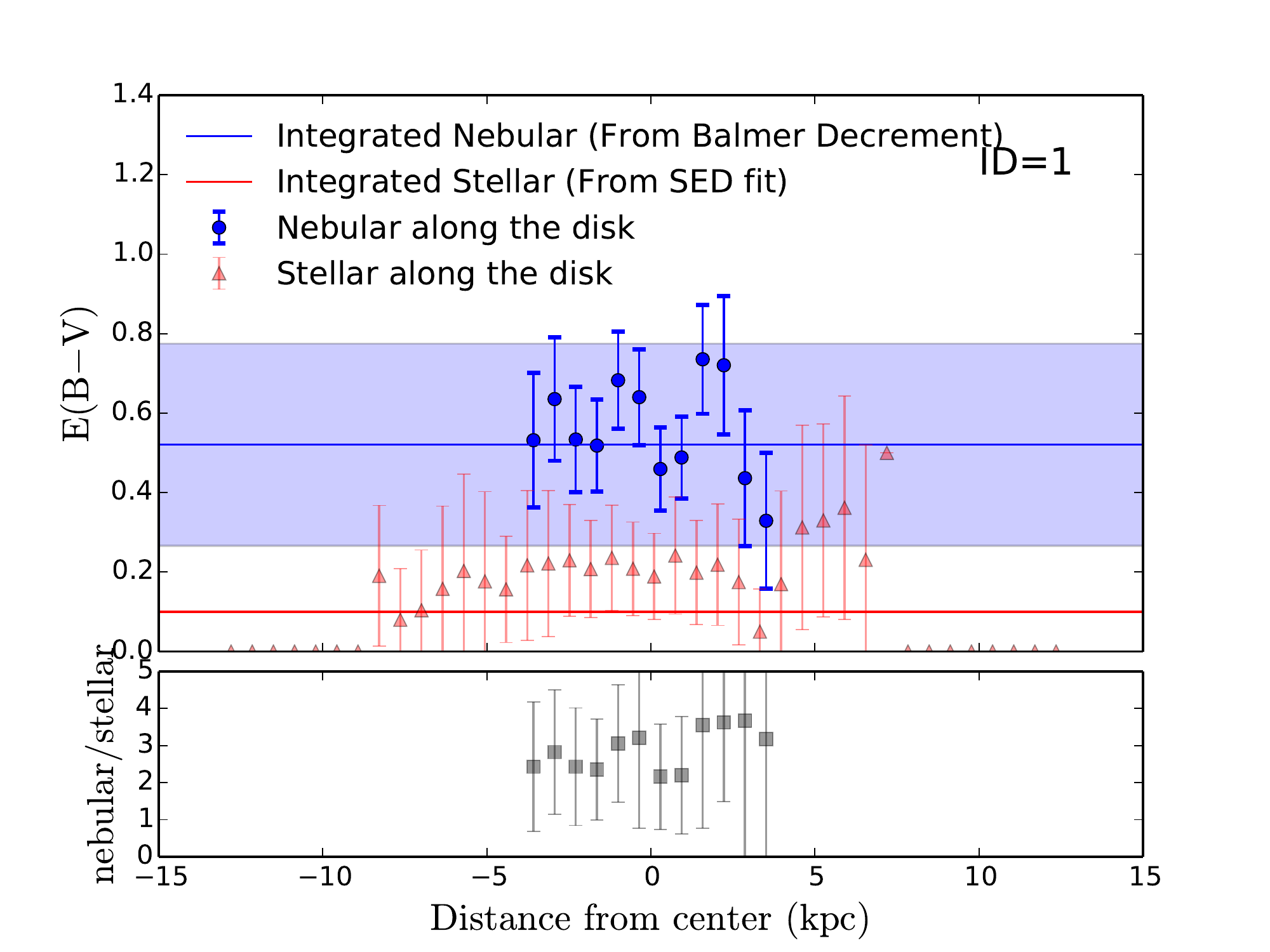}
\includegraphics[trim=0cm 0.1cm 0.5cm 0.5cm, clip,width=7.2cm]{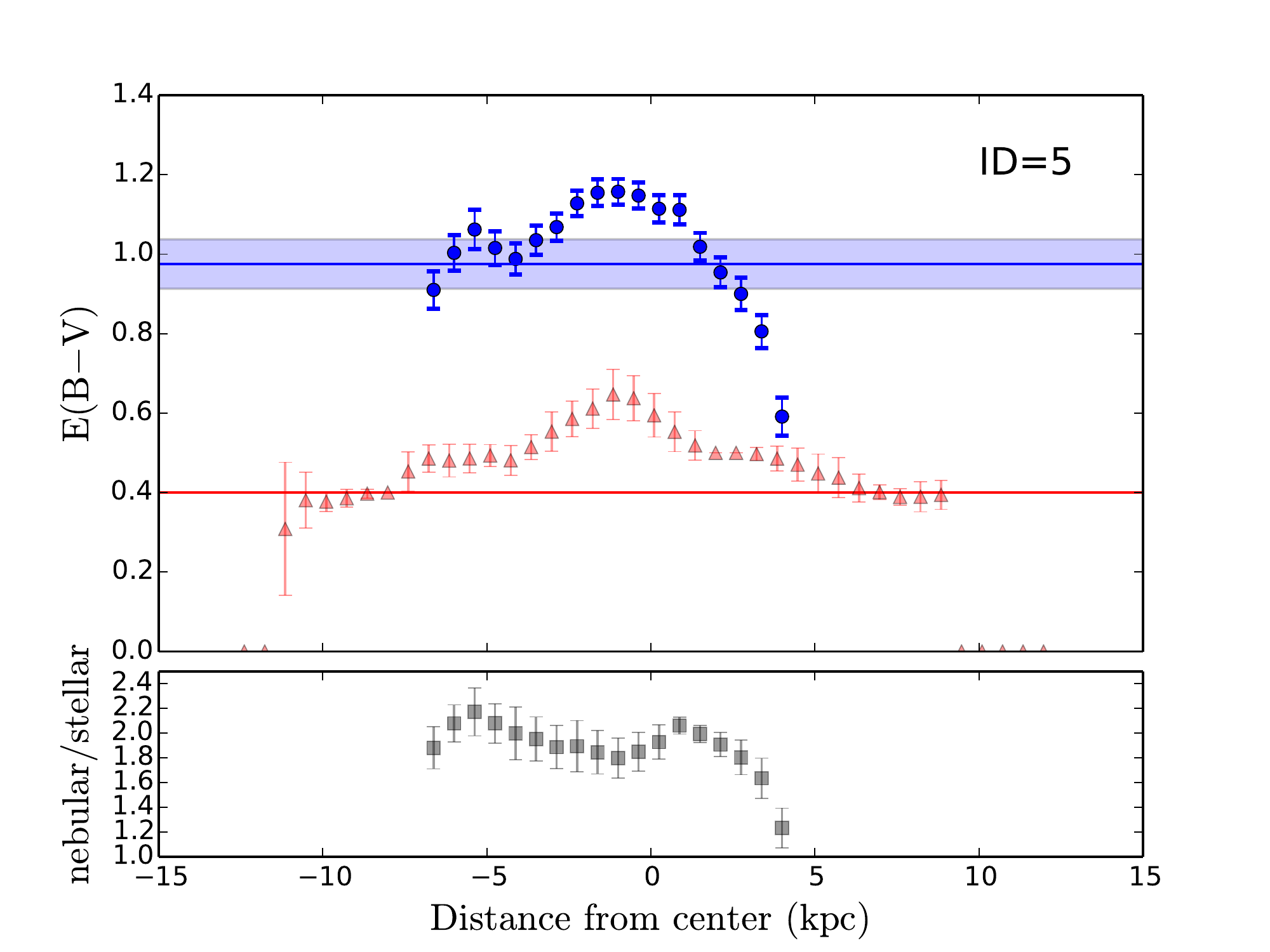}\\
\includegraphics[trim=0cm 0.1cm 0.5cm 0.5cm, clip,width=7.2cm]{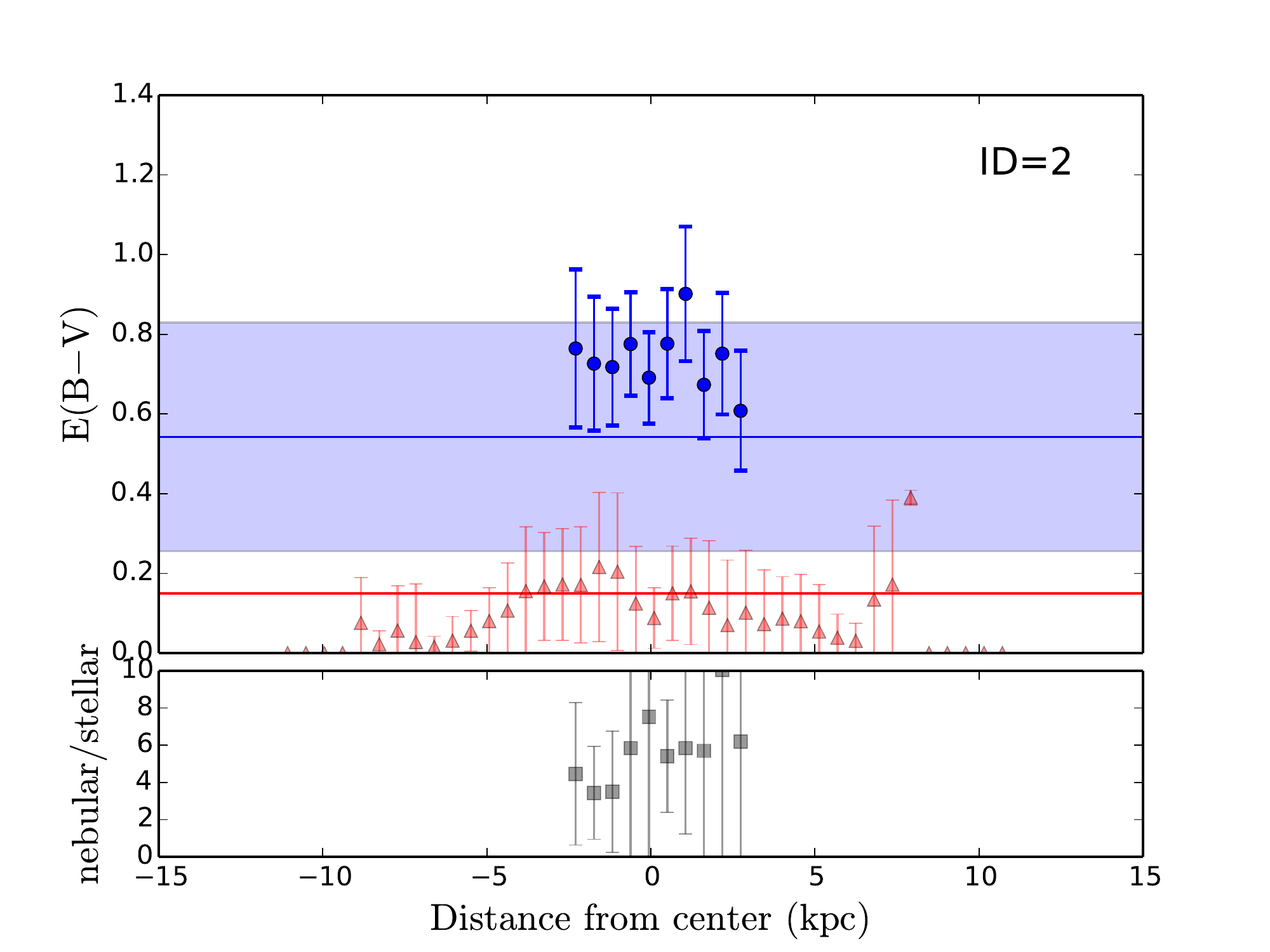}
\includegraphics[trim=0cm 0.1cm 0.5cm 0.5cm, clip,width=7.2cm]{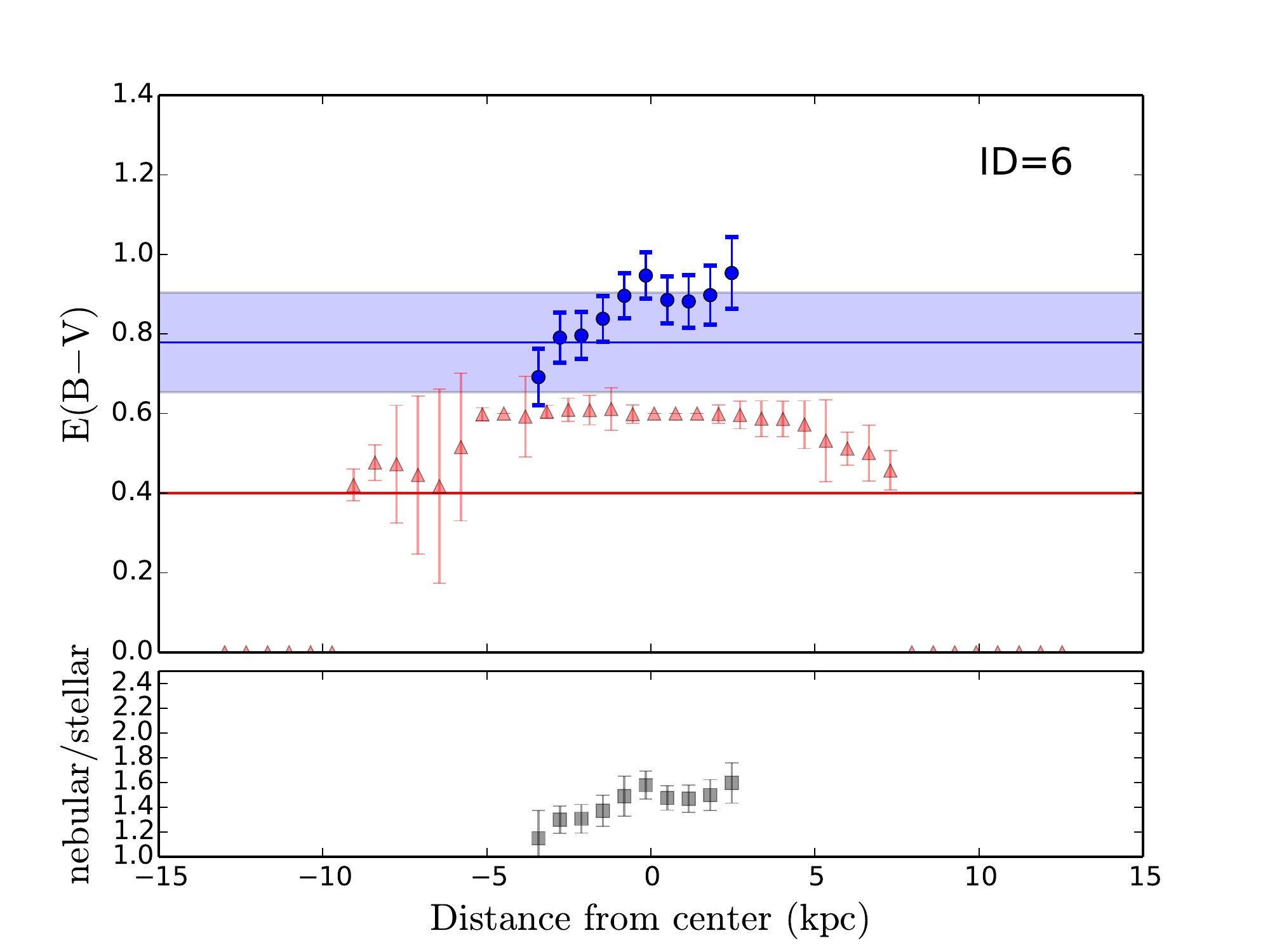}\\
\includegraphics[trim=0cm 0.1cm 0.5cm 0.5cm, clip,width=7.2cm]{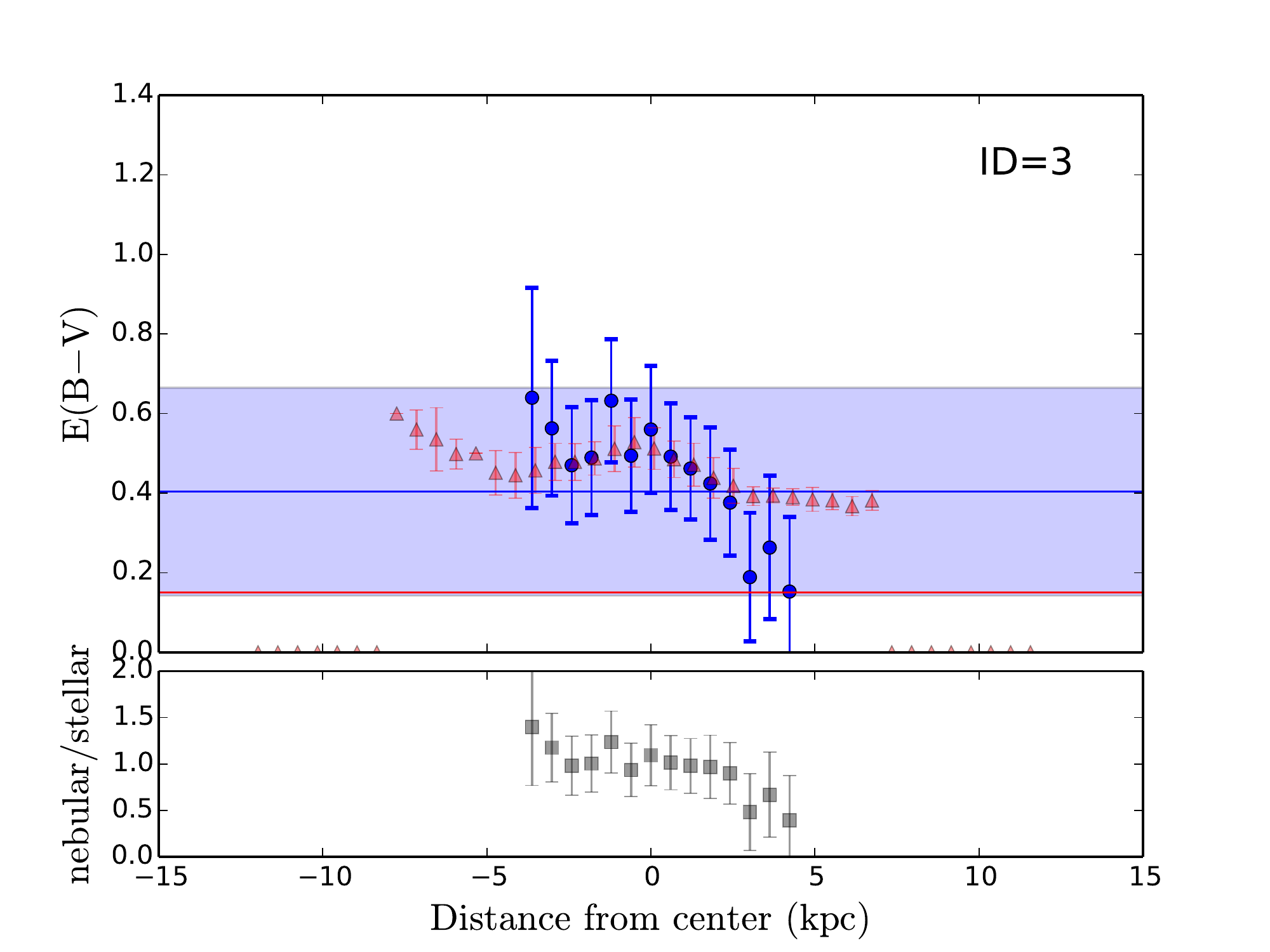}
\includegraphics[trim=0cm 0.1cm 0.5cm 0.5cm, clip,width=7.2cm]{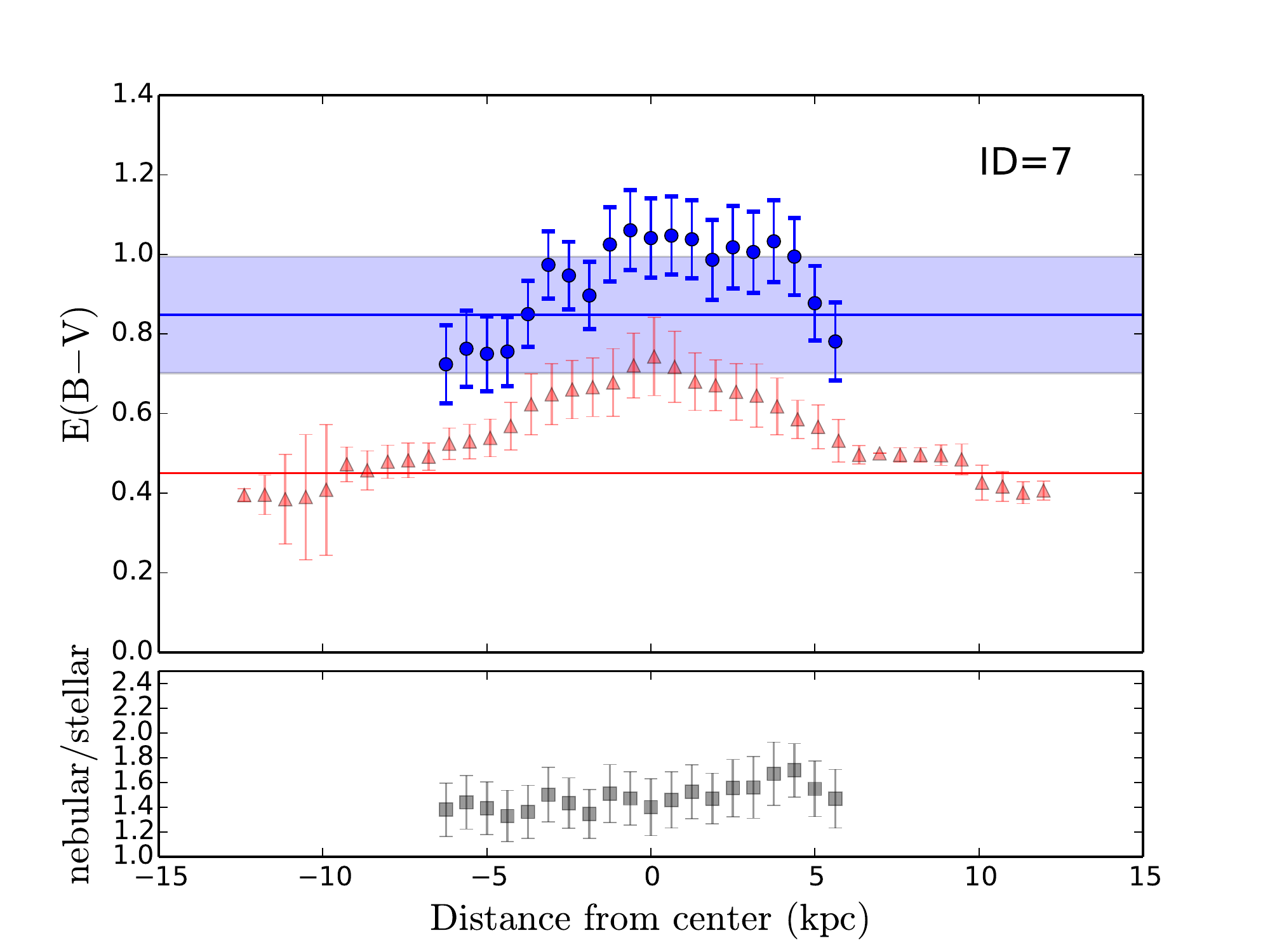}\\
\includegraphics[trim=0cm 0.1cm 0.5cm 0.5cm, clip,width=7.2cm]{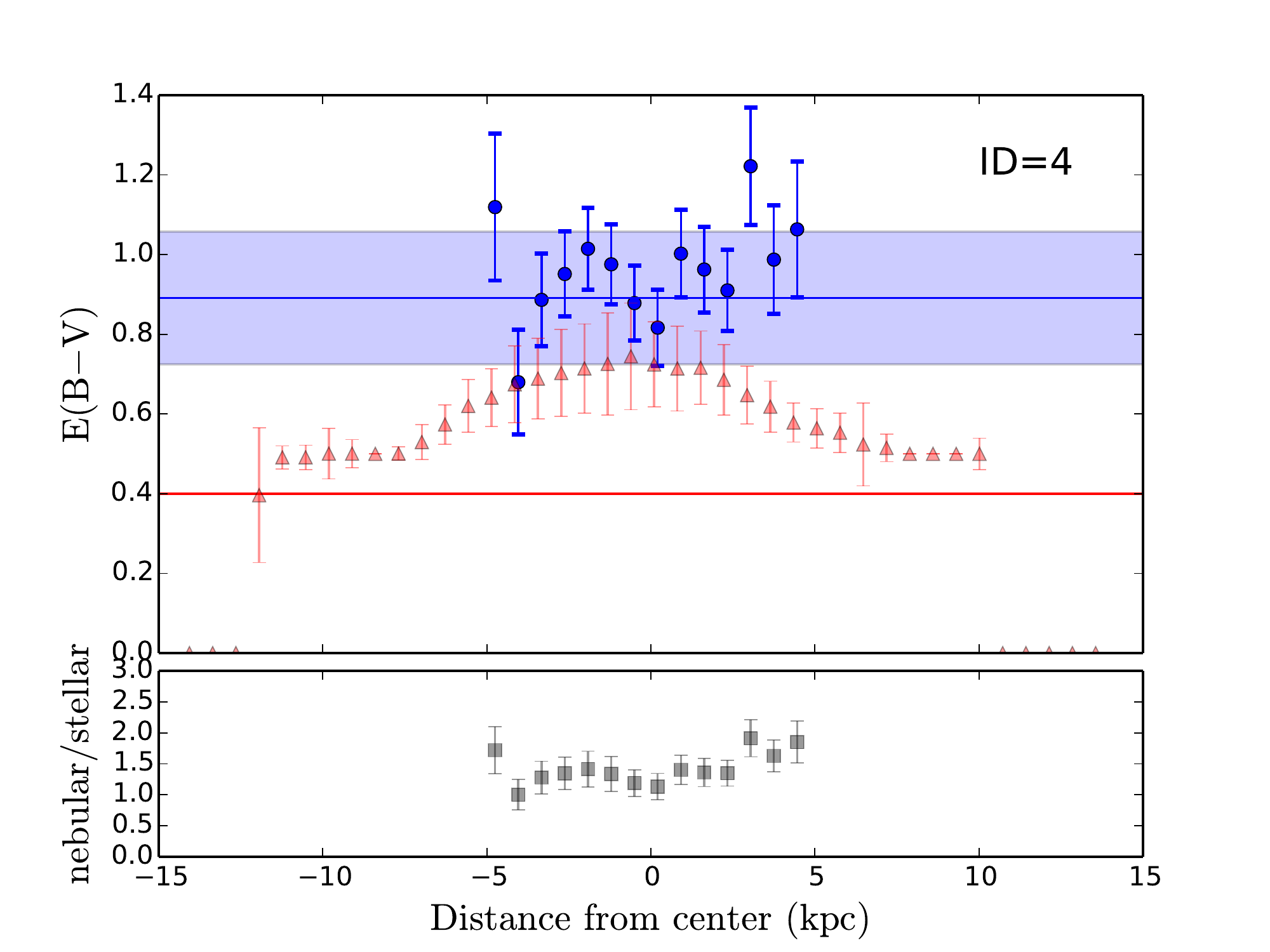}
\includegraphics[trim=0cm 0.1cm 0.5cm 0.5cm, clip,width=7.2cm]{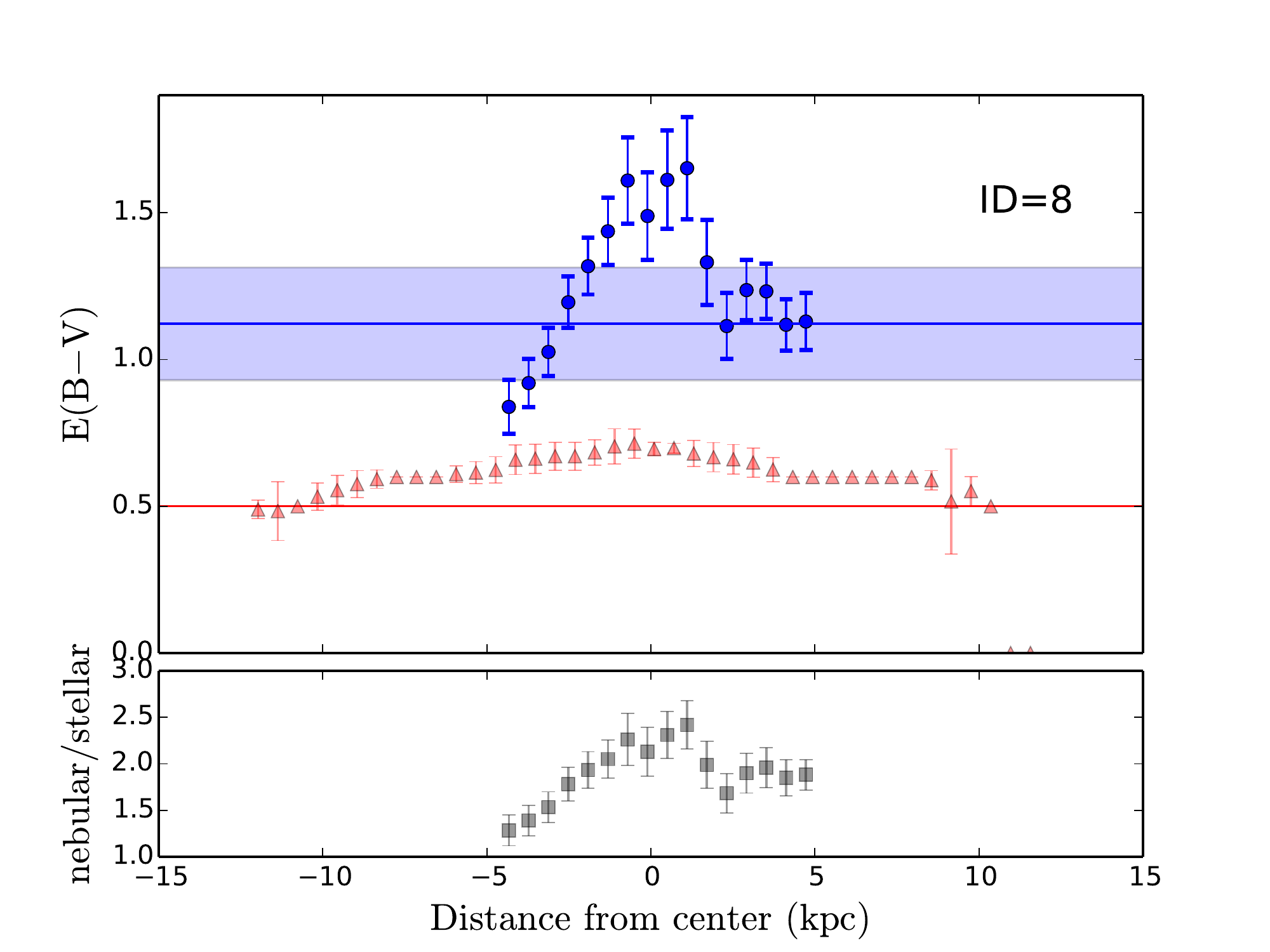}\\
\end{tabular}
\caption{ Nebular and stellar color excess profiles (top panels) and
  nebular to stellar color excess ratios (bottom panels) along the major axis of galaxies in our sample shown with blue circles and
  red triangles respectively as a function of distance from the center
  of galaxies. Blue and red solid lines show the nebular and stellar
  integrated color excess measured in each galaxy and the blue shaded
  region is the $1\sigma$ uncertainty in the integrated nebular color
  excess.}
\end{figure*}

The ratio of the nebular to stellar dust extinction has been studied
extensively over the past decade. Studies of local star forming
galaxies (e.g. \citealt{Calzetti2000}; \citealt{Wild2011}) have found
larger attenuation towards the nebular regions compared to the stellar
continuum. However, in almost all of these works, there is a large
scatter in the nebular vs. stellar color excess relation. Recent works (\citealt {Reddy2015}) have attributed the
scatter in this relation, seen in samples of higher redshift
star-forming galaxies, to physical properties of galaxies,
specifically their stellar mass or sSFR. 

In Figure 4, we compare the integrated nebular and stellar color
excesses of galaxies in our sample. We have color-coded our galaxies
based on their sSFR and the symbol sizes increase with stellar
mass. All our galaxies sit above the $1:1$ (dashed line)
nebular to stellar color excess ratio, which means there is larger
amount of nebular extinction compared to the continuum. While we do
not find any clear trend in this ratio with neither sSFR or stellar
mass, the sample size is too small to draw any strong conclusions.  

We now compare resolved stellar and nebular attenuation measures along the disk of galaxies (as explained in the previous section). Plotted in
Figure 5, are nebular (blue circles) and stellar (red triangles) color
excess as a function of distance (in kpc) from the center of the galaxy, as
well as the global measurement for each galaxy (solid lines). An
agreement is seen among the nebular measurements at the resolution
elements with the global measured value for each galaxy. The difference
between the median of the measured points and the
integrated value in each galaxy ranges from zero to maximum of 0.1
magnitude (well within the uncertainty of the integrated value of each
galaxy). The small offset towards higher median values, might be due to the S/N
criteria which affects the spectra of the outer parts of the disk more
than the central parts. The uncertainties in the measured values
become considerably smaller in galaxies with larger stellar masses (the right panels). Almost all the galaxies show higher
stellar continuum extinction in the central regions compared to the
outer parts in the disk. This differs from the nebular color
excess of the lower mass galaxies ($\rm ID=1-4$) which does not significantly
vary as a function of their distance from the center. More massive
galaxies ($\rm ID=5-8$) however, show higher nebular extinction
towards the central parts of galaxies similar to the stellar continuum
color excesses. 

As shown in Figure 4, the nebular to stellar color
excess ratio, varies from galaxy to galaxy. By examining $\rm ID=5$
and $\rm ID=7$, two galaxies with the most extended emissions, highest S/N
ratio, exact same redshift and comparable stellar masses, we see very
different ionized to continuum color excess ratio. Between the two
galaxies the one with the higher sSFR ($\rm ID=7$) has smaller color
excess ratio, consistent with the findings of \cite{Price2014} for
higher redshift galaxies but inconsistent with \cite{Reddy2015}, and
the framework depicted by radiative transfer models (e.g. \citealt{Charlot2000}).

An important factor that might be playing a role is the
difference in the inclination of these two galaxies. The inclination
affects observed physical properties of galaxies and in particular
their surface brightness (e.g. \citealt{Holmberg1958}). Many studies used inclination
and surface brightness to measure the amount of extinction in disky galaxies
(e.g. \citealt{Giovanelli1995}, \citealt{Peletier1992}), knowing that with
the same surface brightness, the more inclined galaxy suffers more from
dust extinction. More recent study by \cite{Yip2010}, confirmed
the result of higher stellar extinction in more inclined galaxies
using a large sample of local SDSS disk galaxies. More interestingly,
this study showed that the amount of the Balmer decrement stays
constant with inclination. If true, this can explain some of the dispersion
in the ionized vs. stellar dust extinction studies. In the case of
this study, if the more inclined galaxy ($\rm ID=7$) were face-on, it would
have had less stellar continuum color excess and therefore larger ratio
of ionized to stellar color excess similar to the other galaxy ($\rm ID=5$).

We do not find any significant correlation between the
nebular to stellar color excess ratio and distance from the center of
galaxies (see bottom panels of figure 5) except for the most massive galaxy ($\rm ID=8$) in which this
ratio decreases with distance from the center. This is because the
resolution of boxes in which the color excess is measured is still significantly larger compared to the sizes of original clouds (sub-kpc)
producing the recombination lines. However, future telescopes such as
Thirty Meter Telescope (TMT) equipped with integral field unit (IFU)
technology and assisted by adaptive optics (AO) systems will be able
to provide us with a wealth of information on these clouds at these redshifts.

\subsection{Variation of Color Excess with Stellar Mass Surface Density}

The typical dust extinction of a galaxy has been shown to depend upon
different properties of the galaxy, most fundamentally on its stellar
mass (e.g. \citealt{Garn2010}, \citealt{Sobral2012},
\citealt{Ibar2013}). Here, we investigate the variation of color
excess in galaxies as a function of the stellar mass surface density at each resolution
element along the major axis of the disks. 

\begin{figure*}[]
\centering
\begin{tabular}{cc}
\includegraphics[trim=0cm 0cm 0cm 1cm, clip,width=9cm]{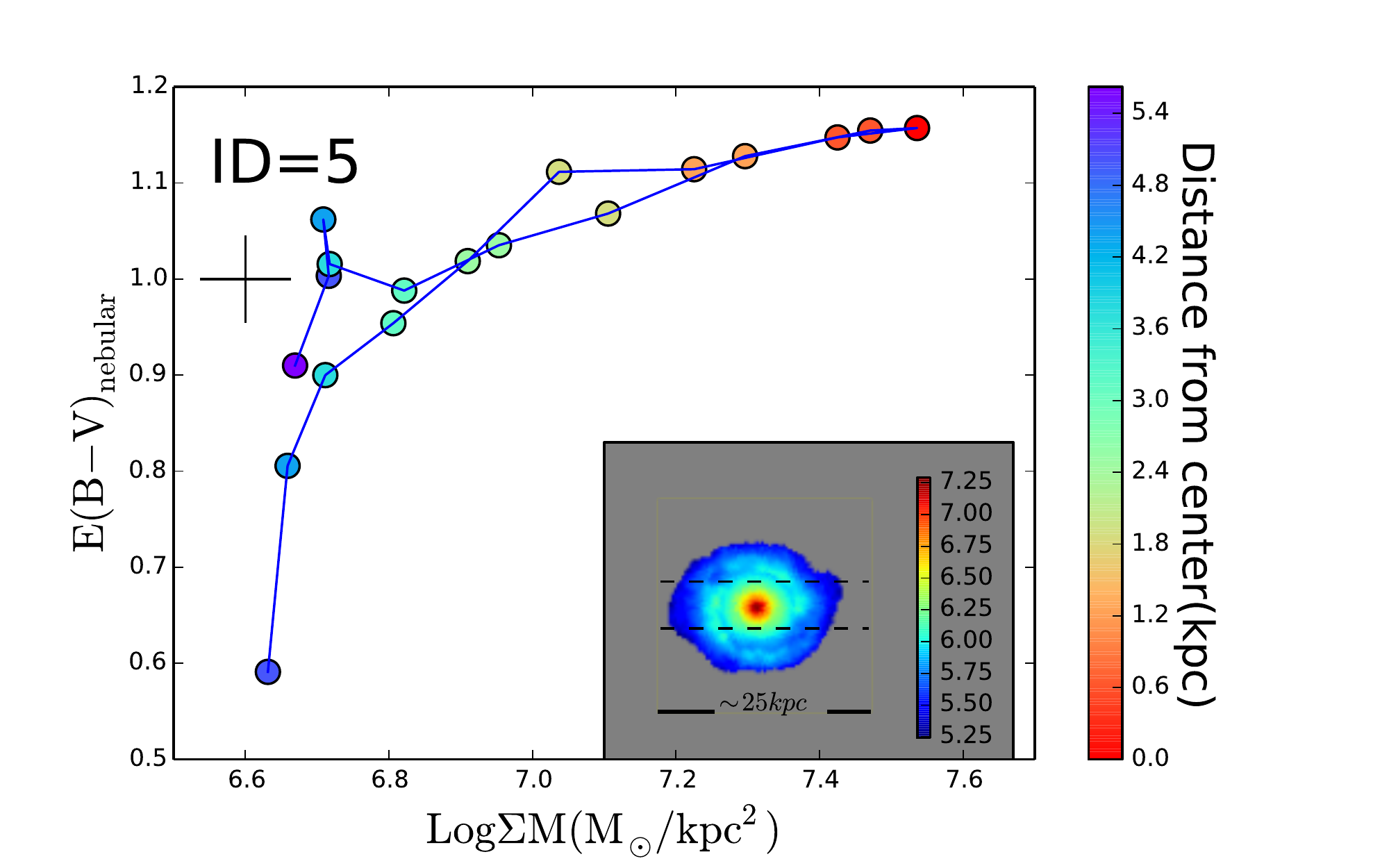}
\includegraphics[trim=0cm 0cm 0cm 1cm, clip,width=9cm]{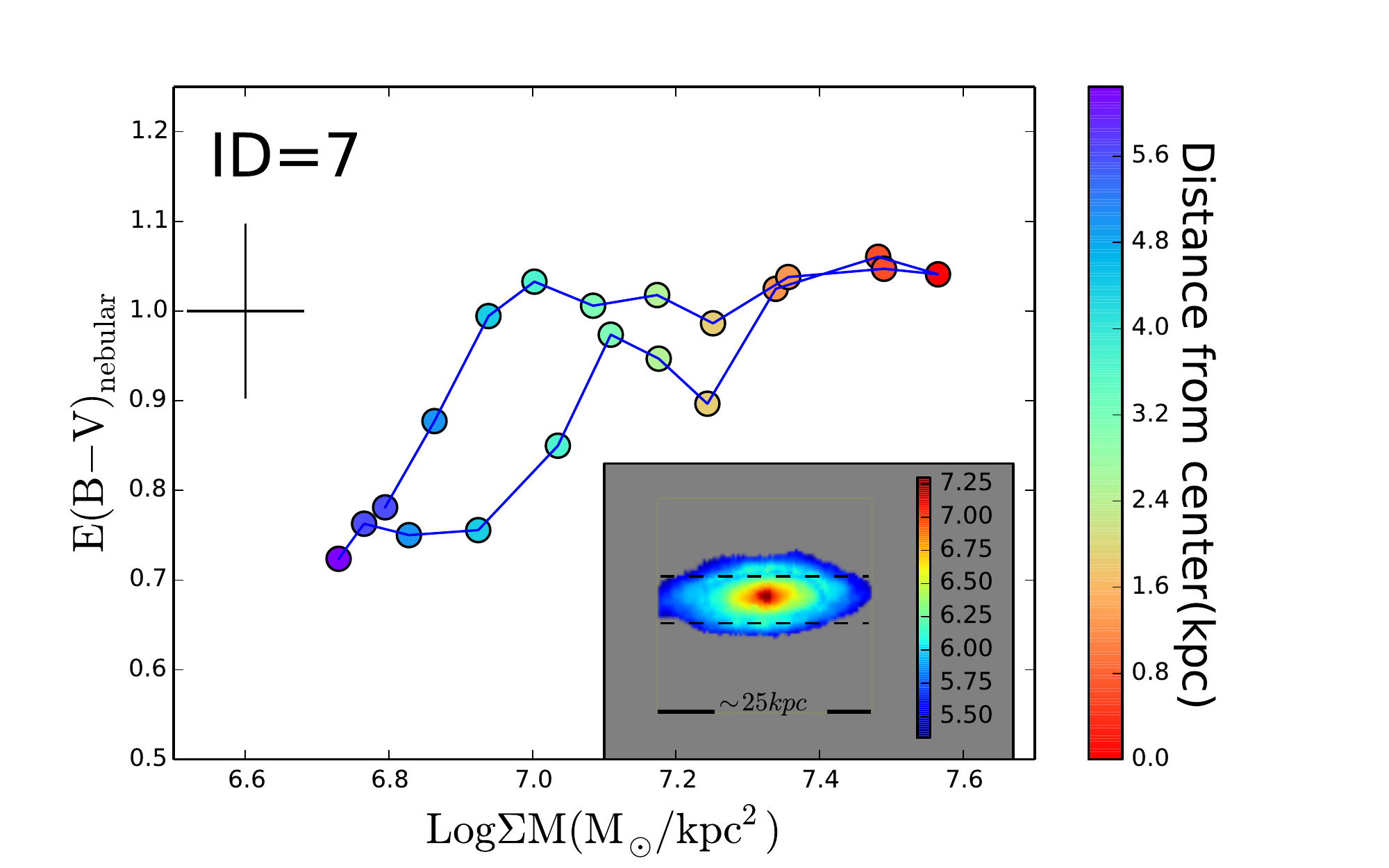}
\end{tabular}
\caption{Variation of nebular color excess as a function of stellar mass
  surface density in two of the galaxies, at the same redshift and
  with the same stellar mass. The stellar mass surface density maps
  are plotted in the corner right of each plot with the DEIMOS slit
  coverage over plotted. The measurements at each resolution element in
  the disk are color coded based on their distance from the center
  (distance from the center decreases going from blue to red color). Average uncertainty in measurements is shown in top left of each panel.}
\end{figure*}

Figure 6, shows the color excess of the ionized gas as a function of stellar mass surface
density for two of the galaxies. The data points are color-coded based on their distance
from the center and the typical error bars are shown in the top-left
corner of each panel. The stellar mass surface density maps from the
resolved SED fitting are shown in the lower right part of each plot
and over plotted with dashed black lines are the DEIMOS slit position
and coverage. It is important to note that while the size of the
disk's major axis is about 20 kpc in these galaxies, the extent of the emission
lines is only about 12 kpc. This can be either due to lack of strong emission
at large distances in these galaxies, or due partly to the over
subtraction of background in larger radii in the DEIMOS reduction
pipeline.

It is clear from figure 6 that the color excess of ionized gas is
increasing from the outskirts towards the bulge of the galaxies
symmetrically, similar to the stellar mass. Putting all the resolution
elements of all galaxies in the sample together on the color excess vs. stellar mass
surface density plot in Figure 7, we see the overall increase of the
median of color excess in bins of stellar mass surface density. 
\begin{figure}[]
\includegraphics[trim=1cm 7cm 1cm 7cm, clip, width=9cm]{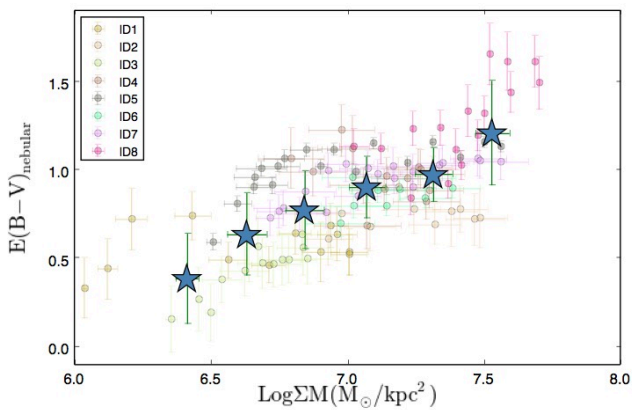}
\caption{Variation of color excess as a function of stellar mass surface density in all resolution elements of the galaxies in the sample. Different colors correspond to different galaxies. There is an overall increasing trend of color excess with increasing stellar mass surface density clear from the median represented with blue stars. }
\end{figure}

\subsection{Dust to Gas Ratio}

A tight correlation exists between the mean surface density of cold gas and
the average SFR per unit area (the so-called SFR law) on
global scales (e.g. \citealt{Kennicutt1998}) as well as on resolved
kpc-scales (\citealt{Kennicutt2007}). This relation (``KS relation'') is parameterized
using a power-law introduced by \cite{Schmidt1963}. Here, we convert
our SFR surface density ($\Sigma_{SFR}$; in units of $\rm
M_{\odot}yr^{-1}kpc^{-2}$) measurements to total gas (molecular and
atomic) surface density ($\Sigma_{H}$; in units of $\rm M_{\odot}pc^{-2}$), using: 
\begin{equation}
\rm log \Sigma_{SFR}=(1.56\pm0.04) log \Sigma_{H} - (4.32\pm0.09)
\end{equation}
derived by \cite{Kennicutt2007} using observations of the nearby spiral
galaxy M51a. We note that as the gas surface densities in higher redshift star
  forming galaxies is larger compared to M51, using the KS relation is an
  extrapolation when dealing with high gas surface densities. There are only limited observations of local galaxies with high
  gas surface densities, and they show evidence of shift in the KS
  relation at higher redshifts \citealt{Hodge2015}. Figure 8 shows the variation of
nebular color excess per resolution elements along the major axis of galaxies in our sample as
a function of gas surface density. There is an increasing trend of
color excess with gas surface density at ($log \Sigma_{H} \gtrsim 2.5
$). We formalize this relation by fitting a third order polynomial to
$E(B-V)_{nebular}$ versus $x \equiv log \Sigma_{H}$ (shown in Figure 8
with magenta dashed line). 
\begin{equation}
E(B-V)_{nebular} = 0.51x^3-3.53 x^2+7.98 x -5.21 
\end{equation}

It is important to note again that, the sample size is small
and we might be missing a population of galaxies that could alter this
fit. In figure 8, we over plot the relation of nebular color
excess as a function of dust mass surface density (multiplied by a
factor of hundred) derived by \cite{Kreckel2013}, from resolved
FIR (Herschel and Spitzer) and optical integral field observations of a
sample of eight nearby disk galaxies. The nebular color excess versus
dust mass surface density relation introduced by \cite{Kreckel2013}
(solid cyan line in Figure 8 ), is well bracketed by the two extremes
of dust geometry discussed in \cite{Calzetti1994} (a uniform foreground dust screen model and a mixed media model),
suggesting a combination of the two effects. The shape of the
color excess versus dust mass surface density resembles that with gas
surface density at higher gas surface
densities. This implies that the change in the dust to gas ratio is insignificant
in these regions with different amounts of extinction, while this is not the case at lower gas
mass surface densities. Many previous studies have assumed a
fixed dust to gas ratio in galaxies (e.g. \citealt{Leroy2011}, \citealt{Sandstrom2013}), this result certifies
the fairness of those assumption at high gas mass surface
densities. The difference in the lower gas mass surface densities,
however is suggestive of a variable gas to dust ratio.

\begin{figure}[]
\includegraphics[trim=2cm 5cm 0cm 4cm, clip, width=10cm]{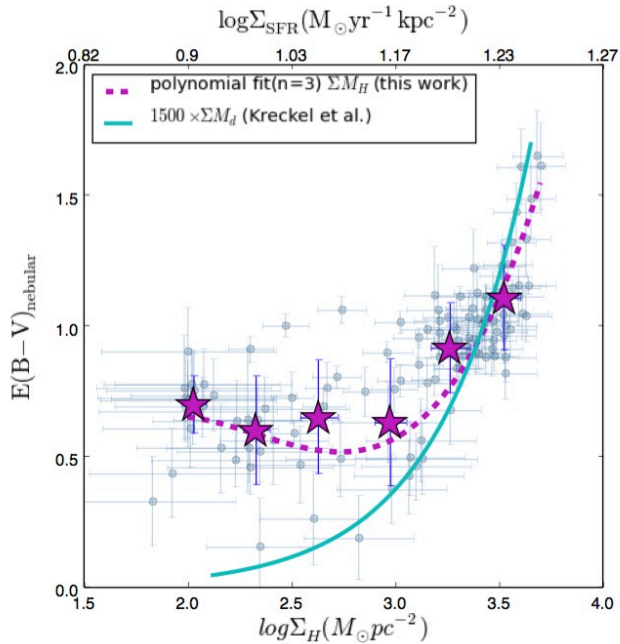}
\caption{Variation of nebular color excess as a function of gas surface
  density in all resolution elements of the galaxies in the
  sample. Magenta stars show the median in
  bins of gas surface density and the magenta dashed line is a third
  order polynomial fit to blue data points. Cyan solid line shows the
  relation between the nebular E(B-V) and dust mass surface density
  (multiplied by a factor of 1500), from \cite{Kreckel2013}.}
\end{figure}
\section{Summary and Discussion}

In this work, we have measured resolved kpc-scale dust reddening along the major axis of
eight emission-line disky galaxies at $z\sim 0.4$. We have used
pixel-by-pixel SED fitting and Keck/DEIMOS spectra to infer stellar
and nebular dust extinction at kpc-scales, respectively. While
the sample size redshift range probed are small to draw robust statistical inferences about
galaxy populations in general, we developed the methodology that can
be practically used on much larger samples to measure dust inside
galaxies at different radii from the center, using optical spectra from local galaxies to $z\sim 0.5$ and infrared spectra at
higher redshifts. These kind of studies could then be used to address
the variation of results for integrated galaxies at high-z found in
the literature.

The integrated nebular to stellar color excess ratio for
the galaxies in our sample are larger than unity with
median $(E(B-V)_{nebular}/E(B-V)_{stellar}) = 2.5$ and standard
deviation $1.0$. Due to the small sample size a clear
trend of nebular to stellar color excess ratio was not seen with either
stellar mass or the sSFR. However the nebular to
stellar color excess ratio dependence can be investigated for
individual galaxies. We specifically compared two of the galaxies in the sample ($\rm ID=5\&7$) with the same redshift
($z=0.38$) and comparable stellar mass ($Log(M_{*}/M_{\odot}\sim
9.65$). Between the two, the one with higher SFR has smaller
nebular to stellar color excess ratio. This is in agreement with the
work of \cite{Price2014} but contrary to the overall trend seen in \citealt{Reddy2015} with integrated
measurements for larger samples of galaxies at higher redshifts. The
difference here can be explained due to differences in the orientations (inclinations) between the two galaxies, a parameter that is often overlooked in many studies.

We also compared the integrated and resolved nebular color excess
in galaxies and found a good agreement between the two, with the
integrated value, being equal or slightly less than the median of
resolved measurements (median $\Delta E(B-V)_{median-integrated} \sim 0.05$ mag). This
small offset is mostly due to the signal to noise criteria applied to
H$\alpha$ and H$\beta$ emission lines at resolution elements, which
exclude the outer parts of the emission line (which also appear to
have lower extinction) from the resolved measurements. We found that
the resolved stellar continuum color excess profiles show higher extinction towards central regions of galaxies compared to outer parts of
the disk. This is contrary the nebular color excess profiles in the
lower mass galaxies which show almost flat radial profiles. 

The relation between the stellar mass and color excess has been
studied at various redshifts (e.g. \citealt{Garn2010},
\citealt{Sobral2012}, \citealt{Dominguez2013}). Here, we extended this relation to
substructures inside galaxies, by studying the variation of nebular color excess as a function of
stellar mass surface density and found an increasing amount of nebular color
excess in regions with higher stellar mass surface density. This also
explains to some extent, the lack of a correlation in the nebular
attenuation as a function of distance from the center in lower mass
galaxies in which the stellar mass surface density range covered is lower, when compared to higher mass galaxies.

We also examined the relation between the nebular color
excess and the gas mass surface densities, converted from the SFR
surface densities. The shape of the attenuation relation for high gas mass surface
densities resembles the attenuation versus dust mass surface density
relation found in \cite{Kreckel2013}, implying that the dust to gas
mass ratio is not changing significantly as a function of extinction
at physical scale probed in this study. It is however important to
note the assumptions made in deriving this result, such as the
conversion from SFR surface density to gas surface density and the
dust mass surface densities which are all measured from local
observations and might not hold true at higher redshifts. Resolved FIR
observations with the Atacama Large Millimeter/submillimeter Array
(ALMA) will be essential to examine the validity of these assumptions. 

\acknowledgments
We thank the anonymous referee for insightful comments which greatly
improved the quality of this manuscript. This work is based on observations at the W.M. Keck Observatory, which
is operated as a scientific partnership among Caltech, the University
of California and NASA.  S.H wishes to thank B. Siana for very
constructive comments and I. Shivaei for useful discussions. DS acknowledges financial support from the Netherlands
Organization for Scientific research (NWO) through a Veni fellowship,
from FCT through a FCT Investigator Starting Grant and Start-up Grant
(IF/01154/2012/CP0189/CT0010) and from FCT grant
PEst-OE/FIS/UI2751/2014.

\bibliography{shooby.bib}

\end{document}